\documentclass[12pt]{emulateapj}
\usepackage{graphicx,rotating}
\usepackage{amssymb}
\usepackage{natbib}
\newcommand{\beq}{\begin{equation}}
\newcommand{\eeq}{\end{equation}}
\newcommand{\etal}{{\sl et~al.~}}
\newcommand{\HST}{{\sl HST~}}
\newcommand{\HIPs}{{\sl Hipparcos~}}

\newcommand{\HIP}{{\sl Hipparcos}}
\newcommand{\Hip}{{\sl Hipparcos}}
\newcommand{\kms}{km s$^{-1}$~}
\def\fdg{\hbox{$.\!\!^\circ$}}

\begin{document}
\bibliographystyle{my2}

\title{Astrometry with the Hubble Space Telescope:
Trigonometric Parallaxes of Selected Hyads\footnote{Based on 
observations made with
the NASA/ESA Hubble Space Telescope, obtained at the Space Telescope
Science Institute, which is operated by the
Association of Universities for Research in Astronomy, Inc., under NASA
contract NAS5-26555}}

\author{ Barbara E.
McArthur\altaffilmark{2}, G.\ Fritz Benedict\altaffilmark{2},    Thomas E. Harrison\altaffilmark{3},    William van Altena\altaffilmark{4}}

\altaffiltext{2}{McDonald Observatory, University of Texas, Austin, TX 78712}
\altaffiltext{3}{Department of Astronomy, New Mexico State University, Las Cruces, New Mexico 88003}
\altaffiltext{4}{ Department of Astronomy, Yale University, New Haven CT 06520 }



\begin{abstract}
We present absolute parallaxes and proper motions for seven members of the Hyades open cluster, pre-selected to lie in the core of the cluster.  Our data come from archival astrometric data from 
FGS 3, and newer data for 3 Hyads from FGS 1R, both white-light interferometers on the {\it Hubble Space Telescope} ({\it HST}). We obtain member parallaxes from six individual Fine Guidance Sensor (FGS) fields and use the field containing van Altena 622 and van Altena 627 (= HIP 21138) as an example. Proper motions, spectral classifications and VJHK photometry of the stars comprising the astrometric reference frames provide spectrophotometric estimates of reference star absolute parallaxes. Introducing these into our model as observations with error, we determine absolute parallaxes for each Hyad. The parallax of vA\,627 is significantly improved by including a perturbation orbit for this previously known spectroscopic binary, now an astrometric binary. Compared to our original (1997) determinations, a combination of new data, updated calibration, and improved analysis lowered the individual parallax errors by an average factor of 4.5. Comparing parallaxes of the four stars contained in the \HIPs catalog, we obtain an average factor of 11 times improvement  with the \HST.  With these new results, we  also have better agreement
 with \HIPs for the four stars in common. These new parallaxes provide an average distance for these seven members, $<{\rm D}>$ = 47.5 pc, for the core a $\pm1-\sigma$ dispersion depth of 3.6 pc, and a minimum depth from individual components of 16.0 $\pm$ 0.9 pc. Absolute magnitudes for each member
are compared to established main sequences, with excellent agreement. 
We obtain a weighted average distance modulus for the core of the Hyades of m-M=3.376 $\pm$ 0.01, a value close to the previous \HIPs values, m-M=3.33$\pm0.02$. 
\end{abstract}




%

\section{Introduction}
What is the value of another parallax for the Hyades open star cluster? Though \cite{Lee09}, Perryman \etal (1998) \nocite{Per98} and \nocite{Bru01} de Bruijne \etal (2001) have established a distance to the Hyades using \HIPs data, there remains a nagging worry; the Pleaides. We \nocite{Sod05} (Soderblom \etal 2005), \cite{Joh05}, and others (Gatewood \etal 2000, Pan \etal 2004, Munari \etal 2004) \nocite{Gat00,Pan04,Mun04} have independently measured a parallax that consistently differs from \HIP, a difference that remains in the recent \HIPs re-reduction  \citep{Lee07a}. Recently \nocite{Pla07} Platais \etal (2007) found a similar distance discrepancy for the young cluster IC 2391. For many of the objects done both by \HST and \HIPs the agreement is good (e.g. Benedict \& McArthur 2005). However, the Hyad parallax values from van Altena \etal (1997) \nocite{WvA97a} were not included in the Benedict \& McArthur (2005) comparison, because it was an initial analysis that pre-dated the improved calibration and Bayesian techniques that we later developed.   Given the importance of the Hyades as a rung in the distance scale ladder \nocite{An07b} (e.g., An \etal 2007b), and the overall importance of \HIPs parallaxes to modern astrophysics, it would be useful to revisit the field and obtain an independent parallax and distance modulus for the Hyades and confirm that, as asserted in \cite{Nar99, Nar99b},  \HIPs got it right.

Our targets are seven confirmed members of the Hyades cluster, with van Altena numbers \nocite{WvA66} (van Altena 1966) vA\,310, 383, 472, 548, 622, 627, and 645. They are distributed on the sky in a roughly circular pattern centered on the cluster center, a consequence of an effort to pick Hyads in the core of the cluster.  As pointed out in \nocite{Per98} Perryman \etal (1998) there were significant differences between the \HIPs parallaxes and the individual
\HST FGS  parallaxes from van Altena \etal (1997). For the three brightest stars
of the four in common between the two sets of observations
(HIP 20563/vA\,310, HIP 20850/vA\,472, HIP 21123/vA\,627) the
\HIPs parallaxes \citep{Lee07a} are between 22--38\% larger than the
1997 \HST values. 

We anticipated that our improved techniques would significantly improve the precision of each Hyad's parallax from the average 1
mas of van Altena \etal (1997). This would potentially translate to a distance precision better than $\sim0.4$pc and an absolute magnitude precision of about 0.02 mag for each Hyades member. In addition to establishing an average parallax for the cluster center,   these seven  individual parallaxes will aid studies of the cluster depth. Our more precise individual absolute magnitudes could better establish the intrinsic width of the main sequence in the Hyades,  identify contaminating binary systems, and provide an independent distance modulus.

In the last eleven years we have substantially improved the process whereby \HST FGS fringe tracking data are turned into parallaxes \nocite{Ben05} (Benedict \& McArthur 2005). This approach has
been applied successfully, resulting in parallax results in many papers
(Benedict \etal 1999, 2000a, 2000b, 2002a, 2002b, 2003, 2006; Beuermann
\etal 2003, 2004; Harrison \etal 1999, 2004; McArthur \etal 1999, 2001, 2010;
Roelofs \etal 2007; 
Soderblom \etal 2005), including parallaxes used for a recent recalibration of the Leavitt Law, the Galactic Cepheid Period-Luminosity relation (Benedict \etal 2007). We report here on applying the improvements to these archival (and newer) FGS data. This effort has resulted in far more accurate and precise  parallaxes for seven Hyads.  \nocite{Ben99, Ben00a, Ben00b, Ben02a, Ben02b, Ben03, Ben06,Ben07,Beu03,Beu04,Har99,Har04,McA99,McA01,McA10,Roe07,Sod05}

Our reduction and analysis of these data is basically the same as for our previous work on galactic Cepheids (Benedict et al. 2007). Our extensive investigation of the astrometric reference stars provides an independent estimation of the line of sight extinction as a function of distance for all reference stars, a significant contributor to the uncertainty in their distances. Using vA\,622$/$627 as an example throughout, we present the results of  spectrophotometry of the astrometric reference stars, information required to derive absolute parallaxes from relative measurements (Section~\ref{SpecPhot});  and derive an absolute parallax for each Hyad (Section~\ref{PIS}). We discuss some astrophysical consequences of these new, more precise distances (primarily the estimation of an independent distance modulus, Section~\ref{HDDM}), and summarize our findings in Section~\ref{SUMRY}.

\cite{Bra91} and \cite{Nel07}
provide an overview of the
FGS instrument and \cite{Ben99}, \cite{Ben02a}, \cite{Har04},  \cite{Ben07} describe the fringe tracking (POS) mode astrometric capabilities 
of an FGS, along with the data acquisition and reduction strategies used in the present study. 
We time-tag all data with a modified Julian Date, $mJD = JD - 2400000.5$.

\section{Observations and Data Reduction}  \label{AstRefs}

We obtained from the \HST archive forty orbits of Guaranteed Time Observation Fine Guidance Sensor (FGS) fringe tracking data secured by the \HST Astrometry Science team using FGS 3. These data contain FGS observations of a total of 7 science targets (confirmed members of the Hyades open cluster listed in Table~\ref{tbl-H}) and 36 reference stars. These data were previously analyzed and resulted in parallaxes for these Hyads published in \cite{WvA97b}. 
We have also secured additional, more recent observations with FGS 1r for three of the Hyads.

Using the vA\,622, vA\,627 field as an example, Figure \ref{fig-1} shows the distribution on the sky of the  Hyads and their reference stars taken  from the Digitized Sky Survey, via {\it Aladin}. For the the vA\,622$/$627 seven sets of astrometric data were acquired with FGS 3 and five sets with FGS 1r the aggregate spanning 16 years, for a total of 164 measurements of vA\,622, 627 and reference stars. Each data set required approximately 33 minutes of spacecraft time. The data were reduced and calibrated as detailed in \cite{Ben02a},  \cite{Ben02b}, \cite{McA01}, and \cite{Ben07}. At each epoch we measured reference stars and the target  multiple times to correct for intra-orbit drift of the type seen in the cross filter calibration data shown in figure 1 of \cite{Ben02a}. 

Table~\ref{tbl-LOO} lists the  epochs of observation for all six of our fields. Ideally ({\it cf.} Benedict et al. 2007) we obtain observations at each of the two maximum parallax factors\footnote{Parallax factors are projections along RA and Dec of the Earth's orbit about the barycenter of the Solar System, normalized to unity.} at two distinct spacecraft roll values imposed by the requirement that
{\it HST} roll to provide thermal control of a camera in the radial bay and to keep its solar panels fully illuminated throughout
the year.  This roll constraint generally imposes alternate orientations at each time of maximum positive or negative parallax factor over a typical two year campaign. A few observations at intermediate or low parallax factors usually allows a clean separation of parallax and proper motion signatures. Unfortunately, we have intermediate observations for only three of our prime targets, vA\,548, vA\,622, and vA\,627. 
For these three fields we were able to take advantage of science instrument command and data handling (SIC\&DH) computer problems that took  the only other then operational science instrument (WFPC2) off-line in late 2008. This situation opened a floodgate  of FGS proposals, temporarily rendering \HST nearly an {\it 'all astrometry, all the time' }mission. Consequently, we obtained additional epochs well-separated in time from the original. This permitted a significantly better determination of relative proper motion for these targets (and for the perturbation orbit of vA\,627, Section~\ref{PO}). For the other Hyad fields two-gyro guiding\footnote{\HST has a full compliment of six rate gyros, two per axis, that provide coarse pointing control. By the time these observations were in progress, three of the gyros had failed. \HST can point with only two. To ``bank" a gyro in anticipation of a future failure, NASA decided to go to two gyro pointing as standard operating procedure.} constraints did not permit re-observation.

\section{Spectrophotometric Absolute Parallaxes of the Astrometric Reference Stars} \label{SpecPhot}
Because the parallax determined for the Hyads will be
measured with respect to reference frame stars which have their own
parallaxes, we must either apply a statistically derived correction from relative to absolute parallax (van Altena, Lee \& Hoffleit 1995, hereafter YPC95 \nocite{WvA95}) or estimate the absolute parallaxes of the reference frame stars listed in Table \ref{tbl-SPP}. In principle, the colors, spectral type, and luminosity class of a star can be used to estimate the absolute magnitude, M$_V$, and V-band absorption, A$_V$. The absolute parallax is then simply,
\beq
\pi_{abs} = 10^{-(V-M_V+5-A_V)/5}
\eeq
The luminosity class is generally more difficult to estimate than the spectral type (temperature class). However, the derived absolute magnitudes are critically dependent on the luminosity class. As a consequence we appeal to reduced proper motions in an attempt to confirm the luminosity classes (see below).

\subsection{Broadband Photometry}
Our band passes for reference star photometry include:  BV (CCD photometry from a 1m telescope at New Mexico State University) and JHK (from 2MASS\footnote{The Two Micron All Sky Survey
is a joint project of the University of Massachusetts and the Infrared Processing
and Analysis Center/California Institute of Technology }). 
Table \ref{tbl-SPP} lists the visible and infrared photometry for all reference stars used in this study.  

\subsection{Spectroscopy, Luminosity Class-sensitive Photometry, and Reduced Proper Motion}
The spectra from which we estimated spectral type and luminosity class come from the New Mexico State University Apache 
Point Observatory\footnote{ The
Apache Point Observatory 3.5 m telescope is owned and operated by
the Astrophysical Research Consortium.}. The dispersion 
was 0.61 \AA/pixel with wavelength coverage 4101 -- 4905 \AA, yielding R$\sim$3700. Classifications used a combination of template matching and line ratios. The brightest targets had about 1500 counts
above sky per pixel, or S/N $\sim$ 40, while the faintest targets had about 400 counts
per pixel (S/N $\sim$ 20). The spectral types for the higher S/N stars are within $\pm$1 subclass. Classifications for the lower S/N stars are $\pm$2
subclasses. Table \ref{tbl-SPP} also lists the spectral types and luminosity classes for our reference stars. 


We  employ the technique of reduced proper motions to provide a confirmation of the reference star estimated luminosity class listed in Table~\ref{tbl-SPP}. We obtain preliminary proper motions ($\mu$)  from UCAC3 \citep{Zac10} and/or PPMXL \citep{Roe10}, and $J$, $K$ photometry from 2MASS for a one-degree-square field centered on each Hyad. With final proper motions from our astrometric solution (Section~\ref{AST}) we plot Figure~\ref{fig-RPM}, which shows $H_K = K + 5\log(\mu)$ versus $(J-K)$ color index for 10,000 stars. If all stars had the same transverse velocities, Figure~\ref{fig-RPM} would be equivalent to an H-R diagram. The Hyads and reference stars are plotted as ID numbers from Table~\ref{tbl-SPP}. Errors in $H_K$, calculated using our final proper motions,  are now $\sim0.3$ mag. 

\subsection{Interstellar Extinction} \label{AV}
To determine interstellar extinction we first plot these stars on several color-color diagrams. A comparison of the relationships between spectral type and intrinsic color against those we measured provides an estimate of reddening. Figure \ref{fig-CCD} contains  a J-K vs V-K color-color diagram and reddening vector for A$_V$ = 1.0. Also plotted are mappings between spectral type and luminosity class V and III from \cite{Bes88} and \cite{Cox00} (hereafter AQ2000). Figure~\ref{fig-CCD}, and similar plots for the other measured colors, along with the estimated spectral types, provides an indication of the reddening for each reference star. 

Assuming an R = 3.1 galactic reddening law (Savage \& Mathis 1979\nocite{Sav79}), we derive A$_V$ values by comparing the measured colors (Table~\ref{tbl-SPP} ) with intrinsic B-V, J-K, and V-K colors from \cite{Bes88} and AQ2000. Specifically we estimate A$_V$ from three different ratios, each derived from the Savage \& Mathis (1979) reddening law: A$_V$/E(J-K) = 5.8; A$_V$/E(V-K) = 1.1; and A$_V$/E(B-V) = 3.1.  The resulting average reference star A$_V$ are collected in Table \ref{tbl-SPP}. 

\subsection{Adopted Reference Frame Absolute Parallaxes}

We derive absolute parallaxes for the reference stars with M$_V$ values from AQ2000 and the $\langle$A$_V\rangle$ derived from the photometry. Our parallax values are listed in Table \ref{tbl-SPP}. We produce errors on the absolute parallaxes by combining contributions from uncertainties in M$_V$ and A$_V$, which we have combined and set to 0.5 magnitude for each reference star. Individually, no reference star parallax is better determined than ${\sigma_{\pi}\over \pi}$ = 23\%.  The average absolute parallax for the vA\,622, 627 reference frame is $\langle\pi_{abs}\rangle = 1.2$ mas.
As a sanity check we compare this to the correction to absolute parallax discussed and presented
in YPC95 (section 3.2, fig. 2). Entering
YPC95, fig. 2, with the vA\,622 galactic
latitude, l = -19\arcdeg, and average magnitude for the
reference frame, $\langle V_{ref} \rangle$= 16.0, we obtain a galactic model-dependent correction
to absolute of 1.3 mas, in agreement.
\section{Absolute Parallaxes of the Hyads}\label{PIS}
Sections 4.1.1-4 detail our astrometric modeling of the vA\,622, 627 field. Any differences in modeling for other Hyads are noted in Section~\ref{ASTnotes}, below. 
\subsection{The vA\,622, 627 Astrometric Model}\label{AST}

 With the positions measured by FGS 3 and FGS 1r we determine the scale, rotation, and offset ``plate
constants" relative to an arbitrarily adopted constraint epoch (the so-called ``master plate") for
each observation set (the data acquired at each epoch). The mJD of each observation set is listed in Table~\ref{tbl-LOO}.
The vA\,622, 627 reference frame contains 6 stars. We employ a four parameter model for those observations. 
For the vA\,622, 627 field all the reference stars have colors similar to the science target. Nonetheless, we also apply the corrections for lateral color discussed in Benedict et al. (1999). 

As for all our previous astrometric analyses, we employ GaussFit \citep{Jef88} to minimize $\chi^2$. The solved equations
of condition for vA\,622, 627 are:
\beq
       x' = x + lc_x(\it B-V) 
\eeq
\beq
        y' = y + lc_y(\it B-V) 
\eeq
\beq
\xi = Ax' + By' + C  - \mu_x \Delta t  - P_\alpha\pi_x
\eeq
\beq
\eta = -Bx' + Ay' + F  - \mu_y \Delta t  - P_\delta\pi_y
\eeq
where $\it x$ and $\it y$ are the measured coordinates from {\it HST};
$\it lc_x$ and $\it lc_y$ are the
lateral color corrections from Benedict et al. 1999\nocite{Ben99}; and $\it B-V $ are
those  colors for each star. A and B  
are scale and rotation plate constants, C and F are
offsets; $\mu_x$ and $\mu_y$ are proper motions; $\Delta$t is the epoch difference from the mean epoch;
$P_\alpha$ and $P_\delta$ are parallax factors;  and $\it \pi_x$ and $\it \pi_y$
 are  the parallaxes in x and y, which are constrained to be equal.   We obtain the parallax factors (projections along RA and Dec of the Earth's orbit about the barycenter of the Solar System normalized to unity) from a JPL Earth orbit predictor \citep{Sta90}, upgraded to version DE405. Additionally, given the previous identification of vA\,627 as a spectroscopic binary \citep{Gri85}, and the higher than typical residuals modeling with only the above equations, we add Kepleran perturbation orbit terms to the model (c.f. \nocite{McA10} McArthur et al. 2010, \nocite{Ben10} Benedict et al. 2010).
 
\subsubsection{Prior Knowledge and Modeling Constraints} \label{MODCON}
In a quasi-Bayesian approach the reference star spectrophotometric absolute parallaxes (Table~\ref{tbl-SPP}) and proper motion estimates for  the reference stars  from 
PPMXL \citep{Roe10} along with the lateral color calibration and B-V color indices were input as observations with associated errors, not as hardwired quantities known to infinite precision.  Input proper motion values  have typical errors of 4--6 mas y$^{-1}$ for each coordinate.  
To assess these input  parallaxes and proper motions, the reference frame is modeled 
without the target to evaluate the goodness of fit of the a priori assumptions.  After the 
target is included   in the modeling, each reference star is sytematically removed one 
at a time to assess impact on the target parallax and proper motions.  Using these techniques we can assess
the inputs for the reference frame, identify double stars in the reference frame, and occasionally  
solve for an orbit for the reference stars that have companions.   Typically, at least 50-100 models are run
in our process to determine the parallax.
We essentially model a 3D volume of the space that contains our science target and reference stars, all at differing distances.

\subsubsection{vA\,627 Perturbation Orbit} \label{PO}
The Keplerian elements for the best-fit perturbation orbit for vA\,627 are presented in Table~\ref{tbl-ORB}. Astrometry from FGS 3 and FGS1r and radial velocities from \cite{Gri85} were modeled simultaneously, using the methods described in \cite{McA10}. The orbit and residuals are presented in Figure~\ref{orb}. Assuming a mass for the K2 V primary, $M_A = 0.74 M_{\sun}$, yields a secondary mass $M_B = 0.42 M_{\sun}$, consistent with an infrared detection of the secondary spectrum by Bender \& Simon (2008)\nocite{Bend08}.  The secondary is evidently an M2 V star.  The estimated magnitude difference between vA\,627 A (K2 V) and vA\,627 B (M2 V) is $\Delta m = 3.6$ magnitudes. The total effect of component B on the apparent magnitude of the vA\,627 system would be $\sim-0.04$ magnitudes.  Hence, the effect of the companion on the size of the actual perturbation orbit (the photocentric orbit, c.f. van de Kamp 1967, Section 11.3) is negligable.\nocite{vdK67}

\subsubsection{Assessing Reference Frame Residuals}
The Optical Field Angle Distortion calibration (McArthur et al. 2002\nocite{McA02}) reduces as-built {\it HST} telescope  and FGS distortions with amplitude $\sim1\arcsec$ to below 2 mas over much of the FGS field of regard. From histograms of the target and reference star astrometric residuals (Figure~\ref{fig-4}) we conclude that we have obtained satisfactory correction in the
region available at all {\it HST} rolls. The resulting reference frame 'catalog' in $\xi$ and $\eta$ standard coordinates (Table \ref{tbl-POS}) was determined, and it has a weighted $<\sigma_\xi>= 0.6$ and $<\sigma_\eta> = 0.7$ mas. Relative proper motions along RA (x) and Dec (y) are also listed in Table \ref{tbl-POS}. The proper motion vector is listed in Table~\ref{tbl-SUM}, as are astrometric results for the other Hyads, including catalog statistics.

To determine if there might be unmodeled - but possibly correctable -  systematic effects at the 1 mas level, we plotted the vA\,622, 627 reference frame x and y residuals against a number of spacecraft, instrumental, and astronomical parameters. These included x,y  position within the pickle-shaped FGS field of regard; radial distance from the center of the FGS field of regard; reference star V magnitude and B-V color; and epoch of observation.  We saw no obvious trends, other than an expected increase in positional uncertainty with reference star magnitude. 

\subsubsection{The Absolute Parallaxes of vA\,622 and vA\,627} \label{AbsPi}
Because of the low ecliptic latitude, most of the parallax signature is along RA.  We obtain for vA\,622 a final absolute parallax $\pi_{abs} = 24.11 \pm0.30$ mas.  This disagrees by almost $3\sigma$ with the \cite{WvA97a} determination, $\pi_{abs} = 21.6 \pm1.1$ mas. We have achieved a significant reduction in formal error. For vA\,627 we obtain  a final absolute parallax $\pi_{abs} = 21.74 \pm0.25$ mas, a value that differs substantially from the  \cite{WvA97a} determination, $\pi_{abs} = 16.5 \pm0.9$ mas. Our new vA\,627 result agrees with previous parallax measurements from \Hip, $\pi_{abs} = 23.4 \pm1.7$ mas \citep{Per98} and $\pi_{abs} = 22.75 \pm1.22$ mas \citep{Lee07a}. We note that this object is another for which the \HIPs re-reduction has improved agreement with \HST. This is not always the case. See \cite{Bar09} for a few counter examples involving galactic Cepheids.
Parallaxes and relative proper motion results for all fields from  {\it HST} and four fields from \HIPs are collected in Table~\ref{tbl-SUM}. Even though \HST both proper motion determinations are relative, the proper motion vectors for  vA\,622, 627 agree  with the absolute motions determined by \HIP. 

\subsubsection{Modeling Notes on the Other Hyads}\label{ASTnotes}
For all targets the reference star average data, \HST  (and if available) \HIPs parallaxes and proper motions are collected in Tables~\ref{tbl-SUM} and \ref{tbl-SUM2}.  In all cases $\it \pi_x$ and $\it \pi_y$ are constrained to be equal.   Three plate models are usually considered  with \HST astrometry.  All models have offset terms C and F.   The differences are in the scale terms.  The first model has an equal scale in $\it x$ and $\it y$, which is the model used for  the vA\,622 and vA\,627 field using equations 4 and 5.    The second model has separate 
scale in $\it x$ and $\it y$,  adding two parameters (D and E) to the first model. 
\beq
\xi = Ax' + By'  +C  - \mu_x \Delta t  - P_\alpha\pi_x
\eeq
\beq
\eta = Dx' + Ey' +F  - \mu_y \Delta t  - P_\delta\pi_y
\eeq
The third model has equal scale in $\it x$ and $\it y$ as the first model does, but also includes  the addition of  two radial terms in each axis (G an H). 
\beq
\xi = Ax' + By' + G(x^2+y^2)+C  - \mu_x \Delta t  - P_\alpha\pi_x
\eeq
\beq
\eta = -Bx' + Ay' + H(x^2+y^2)+F  - \mu_y \Delta t  - P_\delta\pi_y
\eeq
 The number of reference stars and the distribution of those stars dictates the model that is  used.    All fields are tested with all three models and the $\chi^2$ and DOF are compared for goodness of fit.   

{\bf vA\,310} - This field provided six reference stars and we obtained seven usable epochs.  We use a six parameter model, where two terms  (D and E) provide independent scale in y shown in Equations 6 and 7. The \HST parallax, $\pi_{abs} = 20.13 \pm0.17$ mas agrees within the \HIPs errors for both the 1997 and 2007 \HIPs results. Our new parallax is considerably larger than the previous \HST value, $\pi_{abs} = 15.4 \pm0.9$, with a significantly improved formal error.

{\bf vA\,383} -  This field provided eight useful reference stars, but we were only able to secure six useable epochs.  The astrometric model for this field  required the addition of radial terms (G and H), using the 6 parameter model shown in Equations 8 and 9,
The introduction of the radial terms reduced the number of degrees of freedom by 13\%, but reduced the $\chi^2$ by 62\% from the 4 parameter model shown in Equations 4 and 5.  Our vA\,383 parallax is $\pi_{abs} = 21.53 \pm0.20$ mas. The original 1997 \HST value was $\pi_{abs} = 16.0 \pm0.9$ mas.

{\bf vA\,472} -  This field provided four useful reference stars and seven useable epochs. One of the reference stars, ref-86,  is the only giant in our fields, obvious in the reduced proper motion diagram (Figure~\ref{fig-RPM}). In addition to the visual inspection of the classification spectrum and the evidence from the reduced proper motion diagram, a model input that assumes a dwarf classification for ref-86 increases $\chi^2$ by 9\%. The astrometric model for this field is a hybrid using  four parameters (Equations 4 and 5) for one observation set containing an unusable reference star observation, and six parameters (Equations 6 and 7) for the other observation sets. The resulting reference frame 'catalog' in $\xi$ and $\eta$ standard coordinates (Table \ref{tbl-POS}) was determined
with	a weighted $<\sigma_\xi>= 0.9$ and $<\sigma_\eta> = 0.9$ mas. Our vA\,472 parallax is $\pi_{abs} = 21.70 \pm0.15$ mas. This agrees with both the 1997 and 2007 \HIPs results. Our new parallax with a formal error $\sim10$ times smaller disagrees ($4\sigma$) with the van Altena (1997) result.  

{\bf vA\,548} -  Seven reference stars, twelve epochs, and six parameter radial term modeling (Equations 8 and 9) yielded a parallax, $\pi_{abs} = 20.69 \pm0.17$ mas, one that differs substantially from the 1997 \HST value, $\pi_{abs} = 16.8 \pm0.3$ mas.

{\bf vA\,645} -  Five reference stars, six epochs, and six parameter radial modeling (Equations 8, 9) yield $\pi_{abs} = 17.46 \pm0.21$ mas. . 
The vA\,645 parallax agrees  with an average of the 1997 and 2007 \HIPs values. Again, the new \HST result is larger (1.5$\sigma$) than the 1997 \HST parallax.

The parallaxes from the previous analysis of \HST FGS data \citep{WvA97a}; our new analysis, including newer data (Section~\ref{AST});  the original \HIPs results \citep{Per97}; and the recent re-reduction of the \HIPs data \citep{Lee07a, Lee09} are collected in Table~\ref{tbl-pis}.

\subsection{New Analysis Improvements}

In Table ~\ref{tbl-pis} we see that our new analysis yields results that have lower error, are significantly different than our earlier results and in general are more in agreement with both \HIPs results.    Several factors have contributed to this improvement.    We now have a longer baseline on three of our 7 Hyads, and we were able to fit a perturbation orbit to vA\,627.   Our OFAD is greatly improved, with a baseline of 18 years instead of the 3 years of OFAD data we had when the initial Hyades study was done.   Since the early OFAD, which depended upon ground-based proper motions of M35, we have been able to solve for HST-based motions, and we have added additional distortion fitting to  the original OFAD model.   We now have superior information about the reference frame, with improved proper motion and spectrophotometric parallaxes, which we treat as observations with error in the modelling, yielding absolute rather than relative parallaxes.   
All these factors combined yield more accurate and precise results.
The older modelling technique used the  ground based catalog technique of summing 
the reference star information to 0, which is more appropriate for a larger reference frame,
and making adjustments from a relative to absolute parallax.    
The combination of the initial OFAD calibration with the older modelling technique resulted in 
parallaxes that were consistently lower than the new values. The new results are calibrated and modelled consistently with the other HST parallx objects discussed in Section \ref{HEH}.

\subsection{Assessing \HST External Error Using \HIP}\label{HEH}
For the four Hyads in common with \HIP, we obtain an internal parallax precision a factor of eleven better than \HIP. We assess our external accuracy by comparing these and past \HST parallaxes with others from \HIP, specifically the re-reduction of \cite{Lee07a}. A total of twenty-eight stars are listed in Table~\ref{tbl-picomp}, and include exoplanet host stars ($\epsilon$ Eri, $\upsilon$ And, HD 138311, GJ 876, 55 Cnc, HD 38529), binary stars (Wolf 1062 AB, Feige 24, HD 33636, Y Sgr), M dwarfs (Proxima Cen, Barnard's Star, Wolf 1062 AB), Cepheids (l Car, $\zeta$ Gem, $\beta$ Dor, W Sgr, X Sgr, Y Sgr, FF Aql, T Vul, $\delta$ Cep, RT Aur), and the four Hyads of this paper (vA 310, vA 472, vA 627, vA645). We plot \HIP~parallaxes against \HST values in Figure~\ref{fig-HH}. For three of our earliest analyses, rather than utilize spectrophotometrically-derived reference star parallaxes, we applied a model-based correction to absolute parallax discussed in \cite{WvA95}. These are plotted in lighter grey. The regression line is derived from a GaussFit model \citep{Jef88} that fairly assesses errors in both \HST and \HIP~parallaxes. We note no significant scale difference over a parallax range $2 <\pi_{abs} < 770$ mas.

There are few notable outliers in Figure~\ref{fig-HH} and Table~\ref{tbl-picomp}, objects further than $1-\sigma$ from perfect agreement. Most of these are (for \HIP) faint stars. Regarding the two bright Cepheid outliers, RT Aur and Y Sgr, using \HIP~ parallaxes to produce a Period-Luminosity relation  would place RT Aur at least 2.4 mag above the relation (Õat leastÕ because its \HIP~parallax is negative). Y Sgr would lie 1.2 mag below. In this case Cepheid astrophysics supports the accuracy of the \HST parallaxes \citep{Bar09}. Apparently small differences can have significant astrophysical consequences. An \HIP~Pleiades parallax ($\pi_{abs} = 8.32 \pm0.13$ mas, van Leeuwen 2009\nocite{Lee09}) was not included in the Figure~\ref{fig-HH} impartial regression. That value differs only by 0.89 mas from the 2005 \HST~value, yet the equivalent difference of 0.2 magnitude in distance modulus calls much of modern stellar astrophysics into question \citep{Sod05}. The Figure~\ref{fig-HH} regression and errors would predict an \HIPs parallax for the Pleiades $\pi_{abs} = 7.63 \pm0.12$, a $5\sigma$ difference from the measured \HIPs value.

\section{Hyades Depth and Distance Modulus}\label{HDDM}

The high-precision absolute parallaxes in Table~\ref{tbl-pis} (column {\em HST11}) provide us an independent estimate of the depth of the Hyades core. Assuming a Gaussian distribution of Hyads yields a $\pm1-\sigma$ core dispersion of 3.6 pc. Back to front, differencing the distances of vA\,622 and vA\,645 we find a minimum diameter 16.0 $\pm$ 0.9 parsecs. The average distance of this particular sample is D = 47.5 pc.

By computing absolute magnitudes for these seven Hyads we can produce a sparsely populated color - absolute magnitude diagram and estimate a distance modulus, m-M, for the entire cluster.  With parallaxes in hand (Table~\ref{tbl-pis}),  we use vA\,627 as an example to illustrate the steps required to obtain absolute magnitudes for these Hyads. 

\subsection{Absolute Magnitudes and the Lutz-Kelker-Hanson Bias}
When using a trigonometric parallax to estimate the absolute
magnitude of a star, a correction should be made for the
Lutz-Kelker  bias \citep{Lut73} as modified by \cite{Han79}. See \cite{Ben07}, section 5, for a more detailed rationale for the application of this correction to single stars.
Because of the galactic latitude and distance of the Hyades, 
and the scale height of the
stellar population of which it is a member,
we calculate Lutz-Kelker-Hanson (LKH) bias assuming a disk distribution.  The LKH bias is proportional to $(\sigma_{\pi}/\pi)^2$. Presuming that any member of the Hyades belongs to the same class of object as $\delta$ Cep (young Main Sequence stars), we scale the LKH correction determined for $\delta$ Cep in Benedict et al. (2002b) and obtain for vA\,622, LKH = -0.001 magnitude, the maximum correction for any of these Hyads. Thus, LKH bias is a negligible component of the absolute magnitude error budget.

\subsection{The Absolute Magnitude of  vA\,622}
According to \cite{Tay06} Hyades extinction is characterized by E(B-V)$\le0.001$ mag, obviating the necessity for extinction-induced corrections to absolute magnitude or color. 
Adopting for  vA\,622
V= 11.90 $\pm$ 0.01 (SIMBAD)  and  the  absolute parallax, $\pi_{abs} = 24.11 \pm 0.30$ mas from Table \ref{tbl-pis}, we determine a distance modulus, m-M = 3.09$ \pm 0.04$. To obtain a final absolute magnitude, we would normally correct for interstellar extinction. However, with  E(B-V)$\le0.001$ mag, V$_0$ = V = 11.90. The distance modulus and V$_0$ provide for vA\,622 an absolute magnitude M$_V = 8.82 \pm 0.03$. This and the absolute magnitudes for the six  other Hyads are collected in Table~\ref{tbl-AbsM}. All absolute magnitude errors contain only the contribution from the parallax uncertainty.

Figure~\ref{fig-HR} presents an HR diagram constructed from our Hyad absolute magnitudes (Table~\ref{tbl-AbsM}). The figure also contains a Hyades main sequence constructed with V, B-V photometry from \cite{Jon06} transformed to M$_V$ using the \cite{Lee09} distance modulus, and an M67 main sequence from \cite{San04}. There are too few stars with \HST parallaxes to claim any systematic offset from the average Hyades/M67 main sequence.


\subsection{A Hyades Distance Modulus}
Including all seven stars, we obtain a weighted average Hyades distance modulus, m-M = 3.376$\pm$ 0.012. We note that from \HIPs parallaxes both \cite{Per98} and \cite{Lee09} obtain an average distance modulus of m=M=3.33 $\pm$ 0.02 from their entire sample of Hyads. With our entire (small) sample we get a distance modulus that is very close to   \HIPs 1997 or 2007. Our distance modulus determinations are listed in Table~\ref{tbl-mM}, along with other recent distance moduli. We note the agreement between our value, and the results from the studies of binaries yielding orbital parallaxes  \citep{Tor97c, Tor97a, Tor97b}.

Those interested in an even more detailed description of the distance and structure of the Hyades will anticipate the results from {\em Gaia} \citep{Lin08}. Parallax and proper motion precision factors of $10-100\times$ better than \HST are expected by $\sim2018$.

\section{Summary}{\label{SUMRY}

We have reanalyzed older FGS 3 data of six fields and supplemental newer FGS1r astrometric data of two fields in the Hyades, containing seven confirmed Hyads. We employ techniques \citep{Har99, Ben07} devised since the original analysis \citep{WvA97a}. These new absolute parallaxes now provide:
\begin{enumerate}
\item an average distance for these seven members, D = 47.5 pc with individual parallax errors lower by an average factor of 4.5 compared to the original study \citep{WvA97a} and a factor of 11 times better than \HIPs for the four stars in common,

\item a $\pm1-\sigma$ dispersion depth of 3.6 pc, and a minimum diameter 16.0 $\pm$ 0.9 pc,

\item absolute magnitudes for each member, yielding a sparsely populated main sequence,


\item a weighted average distance modulus of m-M=3.376 $\pm$ 0.01, a value that agrees within the errors to results from the orbital parallaxes of Hyades binaries \citep{Tor97c, Tor97a, Tor97b}, and is very close to  both \HIPs results \citep{Per98,Lee09} ,\\

\item an independent parallax and distance modulus for the Hyades confirming the assertion of \cite{Nar99} that \HIPs 'got it right'.
\end{enumerate}

\acknowledgments

Support for this work was provided by NASA through grants NAG5-1603 and AR-11746 from the Space Telescope 
Science Institute, which is operated
by AURA, Inc., under
NASA contract NAS5-26555. These results are based partially on observations obtained with the
Apache Point Observatory 3.5 m telescope, which is owned and operated by
the Astrophysical Research Consortium. This publication makes use of data products from the Two Micron All Sky Survey,
which is a joint project of the University of Massachusetts and the Infrared Processing
and Analysis Center/California Institute of Technology, funded by NASA and the NSF. 
This research has made use of the SIMBAD database and {\it Aladin}, both developed at CDS, Strasbourg, France; the NASA/IPAC Extragalactic Database (NED) which is operated by JPL, California Institute of Technology, under contract with the NASA;  and NASA's Astrophysics Data System Abstract Service. 


\bibliography{myMaster}

\clearpage

\begin{center}
\begin{deluxetable}{llll}
\tablewidth{4in}
\tablecaption{Hyad Positions\label{tbl-H}}
\tablehead{\colhead{vA}&\colhead{alias}&
\colhead{RA~~~~(2000)}&\colhead{Dec}}
\startdata
vA\,310&HIP 20563&04 24 16.94&18 00	10.49 \\
vA\,383&Os 373&04 26 4.71&15 02	28.90 \\
vA\,472&HIP 20850&04 28 04.44&13 52	04.59 \\
vA\,548&BD +15\arcdeg634&04 29 30.98&16	14 41.40 \\
vA\,622&HD 285849&04 31 31.96&17	44	59.10 \\
vA\,627&HIP 21123&04 31 37.10&17	42	35.20 \\
vA\,645&HIP 21138&04 31 52.47&15	29	58.14 \\
\enddata
\end{deluxetable}
\end{center}

\begin{deluxetable}{llll}
\tablewidth{4in}
\tablecaption{Log of Observations\label{tbl-LOO}}
\tablehead{\colhead{Set}&
\colhead{mJD}&\colhead{P$_{\alpha}$\tablenotemark{a}}&\colhead{P$_{\delta}$\tablenotemark{b}} 
}
\startdata
vA\,310 &&&\\
1&49252.9445&0.96275&0.12569\\
2&49407.2201&-1.02765&-0.16291\\
3&49601.8487&1.03632&0.15393\\
4&49768.0651&-1.02281&-0.16692\\
5&49943.0840&1.00037&0.17376\\
6&50128.2371&-1.00929&-0.17027\\
7&50331.9210&1.03261&0.15368\\
vA\,383 &&&\\
2&49409.23137&-1.01368&-0.15877\\
3&49595.01153&1.028972&0.15971\\
4&49781.66987&-1.01164&-0.14403\\
5&49943.95244&0.98764&0.18495\\
6&50123.27825&-0.96919&-0.18435\\
7&50338.95611&0.99744&0.12854\\
vA\,472 &&&\\
1&49256.8956&0.92173&0.08047\\
2&49422.4591&-0.99017&-0.12253\\
3&49596.9344&1.02376&0.15284\\
4&49769.4250&-1.00417&-0.16806\\
5&49963.6348&1.02032&0.14884\\
6&50131.2614&-0.99739&-0.17496\\
7&50340.8985&0.98650&0.11606\\
vA\,548 &&&\\
1&49209.0390&0.97032&0.17966\\
2&49410.1872&-1.02079&-0.15306\\
3&49577.7603&0.98803&0.17667\\
4&49769.3545&-1.01492&-0.16245\\
5&49943.9986&0.99046&0.17542\\
6&50122.2276&-0.96516&-0.17525\\
7&50336.9947&1.01415&0.13361\\
8&54810.5938&-0.17239&-0.11979\\
9&54810.6340&-0.17302&-0.11993\\
10&54811.4594&-0.18770&-0.12178\\
11&54811.5662&-0.18950&-0.12206\\
12&54811.6592&-0.19123&-0.12223\\
vA\,622, vA\,627 &&&\\
1&49227.0396 &1.04440&0.15652\\
2&49408.2456 &-1.02747&-0.15263\\
3&49582.1176 &1.01468&0.16447\\
4&49758.3671 &-0.97649&-0.16396\\
5&49943.8667 &0.99558&0.16578\\
6&50159.1570 &-0.99223&-0.12207\\
7&50331.9914 &1.03557&0.14358\\
8	&55082.2052&	1.03586&	0.14058\\
9	&55086.7759&	1.02332&	0.13315\\
10&	55087.0645	&1.02224&	0.13269\\
11&	55087.5746	&1.02047&	0.13177\\
12	&55089.5716	&1.01252	&0.12822\\
vA\,645 &&&\\
1&49410.30146&-1.01741&-0.15020\\
2&49582.04762&1.00243&0.17306\\
3&49770.49773&-1.01316&-0.15934\\
4&49944.06818&0.98478&0.17634\\
5&50159.96080&-0.97731&-0.10778\\
6&50339.02691&1.00585&0.12372\\
\enddata
\tablenotetext{a}{Parallax factor in Right Ascension }
\tablenotetext{b}{Parallax factor in Declination}
\end{deluxetable}

\begin{deluxetable}{lllllllll}
\tablewidth{0in}
\tablecaption{Hyad Astrometric Reference Star Photometry, Spectral Classifications, and
Estimated Spectrophotometric Parallaxes \label{tbl-SPP}}
\tablehead{\colhead{ID}& \colhead{V}&
\colhead{B-V} & \colhead{V-K} &  \colhead{J-K} &\colhead{SpT} & \colhead{M$_V$} & \colhead{A$_V$} &
\colhead{$\pi_{abs}$}(mas) } 
\startdata
vA\,310\tablenotemark{a} 		&10.01	&	1.02	&		&		&		&		&		&	\\
50	&	15.07	&	1.25	&	2.97	&	0.66	&	K1V	&	6.2	&	1.1	&	2.8$\pm$0.7\\
51	&	15.69	&	0.94	&	2.33	&	0.49	&	G1V	&	4.5	&	1.0	&	0.9$\pm$0.2\\
52	&	14.67	&	1.33	&	3.29	&	0.77	&	K1V	&	6.2	&	1.4	&	 3.8$\pm$0.9\\
53	&	15.57	&	0.84	&	2.11	&	0.42	&	F4V	&	3.3	&	1.3	&	0.7$\pm$0.2\\
54	&	16.11	&	1.23	&	2.86	&	0.65	&	K1V	&	6.2	&	1.1	&	 1.7$\pm$0.4\\
55	&	15.40	&	0.85	&	1.98	&	0.40	&	F4V	&	3.3	&	1.4	&	0.7$\pm$0.2\\
vA\,383 		&12.2	&	1.45	&		&		&		&		&		&	\\
61	&	16.14	&	1.14	&	2.88	&	0.61	&	G2V	&	4.7	&	1.6	&	 1.1$\pm$0.2\\
62	&	15.43	&	1.25	&	3.22	&	0.76	&	G7V	&	5.4	&	1.6	&	 2.1$\pm$0.5\\
63	&	15.06	&	1.1	&	2.87	&	0.65	&	F5V	&	3.5	&	2.0	&	 1.3$\pm$0.3\\
64	&	14.13	&	1.18	&	2.99	&	0.64	&	G3V	&	4.8	&	1.7	&	 3.0$\pm$0.7\\
65	&	15.97	&	1.47	&	2.86	&	0.80	&	K1.5V	&	6.3	&	1.7	&	 2.6$\pm$0.6\\
66	&	16.54	&	1.38	&	2.96	&	0.78	&	K1V	&	6.2	&	1.6	&	1.7$\pm$0.4\\
68	&	16.87	&	1.35	&	4.21	&	0.75	&	G7V	&	5.4	&	1.9	&	 1.2$\pm$0.3\\
69	&	15.24	&	1.14	&	2.58	&	0.58	&	F3V	&	3.2	&	2.3	&	 1.1$\pm$0.3\\
vA\,472 	&	9.1	&	0.8	&		&		&		&		&		&	\\
86	&	14.07	&	1.46	&	3.77	&	0.88	&	K1III	&	0.6	&	1.8	&	0.4$\pm$1.4\\
88	&	15.10	&	1.51	&	3.79	&	0.86	&	K3V	&	6.8	&	1.5	&	4.3$\pm$1.0\\
89	&	16.27	&	0.95	&	2.75	&	0.61	&	G0V	&	4.2	&	1.5	&	0.7$\pm$0.2\\
92	&	17.03	&	1.36	&	5.05	&	0.84	&	M2V	&	9.9	&	1.0	&	 3.1$\pm$0.7\\
vA\,548 &		10.32	&	1.16	&		&		&		&		&		&	\\
96	&	13.44	&	1.07	&	2.65	&	0.57	&	G2V	&	4.7	&	1.4	&	 3.4$\pm$0.8\\
97	&	15.37	&	0.98	&	2.77	&	0.60	&	G0V	&	4.4	&	1.2	&	1.1$\pm$0.3\\
98	&	15.56	&	1.19	&	3.15	&	0.68	&	G1V	&	4.5	&	1.8	&	 1.5$\pm$0.3\\
99	&	15.93	&	1.15	&	3.25	&	0.66	&	G2V	&	4.7	&	1.6	&	 1.2$\pm$0.3\\
100	&	14.05	&	0.95	&	2.55	&	0.46	&	F3V	&	3.2	&	1.7	&	 1.5$\pm$0.3\\
101	&	14.07	&	0.96	&	2.61	&	0.47	&	F4V	&	3.3	&	1.7	&	 1.6$\pm$0.4\\
102	&	14.91	&	1.15	&	3.02	&	0.60	&	G3V	&	4.8	&	1.6	&	 2.0$\pm$0.5\\
vA\,622 	&	11.9	&	1.41	&		&		&		&		&		&	\\
vA\,627 	&	9.53	&	0.99	&		&		&		&		&		&	\\
109	&	14.79	&	0.89	&	2.33	&	0.50	&	G0V	&	4.4	&	1.0	&	 1.3$\pm$0.3\\
112	&	15.64	&	1	&	2.45	&	0.54	&	G2V	&	4.7	&	1.2	&	 1.1$\pm$0.3\\
113	&	13.97	&	0.8	&	2.23	&	0.45	&	F4V	&	3.3	&	1.2	&	 1.3$\pm$0.3\\
115	&	16.10	&	0.89	&	2.44	&	0.51	&	F4V	&	3.3	&	1.5	&	0.6$\pm$0.1\\
116	&	15.46	&	0.89	&	2.43	&	0.52	&	G1V	&	4.5	&	0.9	&	 1.0$\pm$0.2\\
117	&	14.55	&	1	&	2.68	&	0.59	&	G5V	&	5.1	&	1.0	&	 2.0$\pm$0.5\\
vA\,645 	&	11	&	1.21	&		&		&		&		&		&	\\
120	&	15.88	&	1.29	&	3.64	&	0.72	&	F7V	&	3.9	&	2.4	&	 1.2$\pm$0.3\\
121	&	15.28	&	1.15	&	3.12	&	0.61	&	F4V	&	3.3	&	2.3	&	 1.2$\pm$0.3\\
122	&	16.23	&	1.55	&	4.13	&	0.85	&	G2V	&	4.7	&	2.9	&	 1.8$\pm$0.4\\
123	&	16.38	&	1.21	&	3.79	&	0.81	&	G5V	&	5.1	&	1.6	&	 1.2$\pm$0.3\\
126	&	15.98	&	1.3	&	3.60	&	0.70	&	G1.5V	&	4.6	&	2.2	&	 1.5$\pm$0.3\\
\enddata
\tablenotetext{a}{V, B-V for all but vA\,645 from co-I Harrison (NMSU)}
\end{deluxetable}

\clearpage
\begin{deluxetable}{lrrrrr}
\tablewidth{6.0in}
\tablecaption{vA\,622,  vA\,627, and Reference Star Astrometric Data    \label{tbl-POS}}
\tablehead{\colhead{ID}&
\colhead{$\xi$ \tablenotemark{a}} &
\colhead{$\eta$ \tablenotemark{a}} &
\colhead{$\mu_x$ \tablenotemark{b}} &
\colhead{$\mu_y$ \tablenotemark{b}}&\colhead{$\pi_{abs}$\tablenotemark{c}} }
\startdata
vA\,622 \tablenotemark{d}&-22.3540$\pm$0.0003&69.9153$\pm$0.0004&102.21$\pm$0.04&-32.60$\pm$0.03&24.11$\pm$0.30\\
vA\,627&86.5143$\pm$0.0003&21.4862$\pm$0.0003&105.87$\pm$0.04&-30.86$\pm$0.04&21.74$\pm$0.25\\
113&-231.3496 $\pm$0.0005&-40.4306$\pm$0.0005&0.53$\pm$0.06&-2.94$\pm$0.09&1.30$\pm$0.14\\
112&-283.1685$\pm$0.0006&-20.8358$\pm$0.0008&0.52$\pm$1.48&0.38$\pm$1.21&0.57$\pm$0.07\\
116&-300.6054$\pm$0.0011&-2.0963$\pm$0.0011&6.23$\pm$0.06&-5.77$\pm$0.07&1.11$\pm$0.12\\
109&-198.8201$\pm$0.0006&18.9788$\pm$0.0012&6.59$\pm$0.05&-8.81$\pm$0.05&1.29$\pm$0.14\\
111&234.1750$\pm$0.0024&-43.6231$\pm$0.0021&2.66 $\pm$1.12&-2.54$\pm$1.14&0.56$\pm$0.06\\
115&276.3116$\pm$0.0015&35.7347$\pm$0.0019&3.08$\pm$0.62&-3.0$\pm$0.72&0.98$\pm$0.11\\
117&340.4222$\pm$0.0015&2.2211$\pm$0.0009&15.67$\pm$0.67&-16.64$\pm$0.55&1.96$\pm$0.24\\
\enddata
\tablenotetext{a}{$\xi$ (RA) and $\eta$ (Dec) are relative positions in arcseconds. 
}
\tablenotetext{b}{$\mu_x$ and $\mu_y$ are relative motions in mas yr$^{-1}$, where $x$ and $y$ are aligned with RA and Dec. }
\tablenotetext{c}{Absolute parallax in mas}
\tablenotetext{d}{RA = 4$^h$31$^m$31.96$^s$ +17\arcdeg 44' 59\farcs1, J2000, epoch = mJD
50331.9705}
\end{deluxetable}
\clearpage
\begin{center}
\begin{deluxetable}{ll}
\tablecaption{ Orbital Elements of ~Perturbation Due to vA\,627 B  \label{tbl-ORB}}
\tablewidth{0in}
\tablehead{\colhead{Parameter} &  \colhead{Value}
}
\startdata
$\alpha_A$&14.58 $\pm$ 0.24 mas\\
P& 843.94 $\pm$ 0.34 d\\
P& 2.31 $\pm$ 0.001 yr\\
T$_0$ & 48658 $\pm$4 mJD \\
e& 0.20 $\pm$ 0.01\\
i& 134\fdg1 $\pm$ 0\fdg9 \\
$\Omega$&251\arcdeg $\pm$ 4\arcdeg \\
$\omega$&334\arcdeg $\pm$ 2\arcdeg \\
$K_1$&6.34 $\pm$0.3 \kms \\
$M_A$ & 0.83$\pm$0.05$M_{\sun}$\\
$M_B$ & 0.42$\pm$0.05$M_{\sun}$\\
\enddata
\end{deluxetable}
\end{center}

\begin{center}
\begin{deluxetable}{rlll}
\tablecaption{Reference Frame Statistics and Hyad Parallax and Proper Motion\label{tbl-SUM}} 
\tablewidth{7in}
\tablecolumns{8} 
\tablehead{\colhead{Parameter} & \multicolumn{3}{c}{vA}}
\startdata
Hyad & 310&383&472\\
\HST Study Duration (y)&2.95&2.54&2.97\\
Observation Sets (\#)&7&6&7\\
Ref stars (\#)&6&8&3\\
Ref stars $\langle$V$\rangle$&15.42&15.55&14.59\\
Ref stars $\langle$B-V$\rangle$&1.07&1.22&1.49\\
$<\sigma_\xi>$ (mas)   & 0.3 & 0.4 & 0.9 \\
$<\sigma_\eta>$ (mas)& 0.4 & 0.4 & 0.9 \\
{\it HST} $\pi_{abs}$ (mas)&20.13$\pm$0.17&21.53$\pm$0.20&21.70$\pm$0.15\\
{\it HST} Relative $\mu$ (mas y$^{-1}$)&114.44$\pm$0.27&102.60$\pm$0.32&104.69$\pm$0.21\\
in Position Angle (\arcdeg)&106.1$\pm$0.1&103.3$\pm$0.1&104.66$\pm$0.05\\
\\
{\it HIP97} $\pi_{abs}$ (mas)&19.39$\pm$1.79&&21.29$\pm$1.91\\
{\it HIP97}  $\mu$ (mas y$^{-1}$)&117.51$\pm$3.18&&108.70$\pm$1.96\\
in Position Angle (\arcdeg)&106.4&&102.4\\
{\it HIP07} $\pi_{abs}$ (mas)&19.31$\pm$1.93&&21.86$\pm$1.71\\
{\it HIP07}  $\mu$ (mas y$^{-1}$)&116.05$\pm$3.52&&108.19$\pm$1.62\\
in Position Angle (\arcdeg)&107.7&&99.5\\
\enddata
\end{deluxetable}
\end{center}

\begin{center}
\begin{deluxetable}{rllll}
\tablecaption{Reference Frame Statistics and Hyad Parallax and Proper Motion\label{tbl-SUM2}} 
\tablewidth{7in}
\tablecolumns{8} 
\tablehead{\colhead{Parameter} & \multicolumn{4}{c}{vA}}
\startdata
Hyad & 548&622&627&645\\
\HST Study Duration (y)&15.34&16.05&16.05&2.05\\
Observation Sets (\#)&12&12&12&6\\
Ref stars (\#)&7& 6&6 &5\\
Ref stars $\langle$V$\rangle$&14.74&15.09&15.09&15.95\\
Ref stars $\langle$B-V$\rangle$&1.09&0.91&0.91&1.30\\
$<\sigma_\xi>$ (mas)   &  0.4 & 0.6 & 0.6 &0.5\\
$<\sigma_\eta>$ (mas)&  0.3 & 0.7 & 0.7 &0.6\\
{\it HST} $\pi_{abs}$ (mas)&20.69$\pm$0.17&24.11$\pm$0.30&21.74$\pm$0.25&17.46$\pm$0.21\\
{\it HST} Relative $\mu$ (mas y$^{-1}$)&105.74$\pm$0.03&107.28$\pm$0.05&110.28$\pm$0.05&101.81$\pm$0.76\\\
in Position Angle (\arcdeg)&101.18$\pm$0.02&107.7$\pm$0.1&106.3$\pm$0.1&105.19$\pm$0.35\\
\\
{\it HIP97}  $\mu$ (mas y$^{-1}$)&&&110.25$\pm$2.0&103.44$\pm$5.6\\
in Position Angle (\arcdeg)&&&106.3&103.1\\
{\it HIP07} $\pi_{abs}$ (mas)&&&22.75$\pm$1.22&19.13$\pm$5.45\\
{\it HIP07}  $\mu$ (mas y$^{-1}$)&&&110.23$\pm$1.36&100.91$\pm$4.3\\
in Position Angle (\arcdeg)&&&105.9&104.2\\
\enddata
\end{deluxetable}
\end{center}

\begin{deluxetable}{cccccccc}
\tablewidth{0in}
\tablecaption{Hyad Absolute Parallaxes (mas)\label{tbl-pis}}
\tablehead{\colhead{vA}& \colhead{HST(97)\tablenotemark{a}}&
\colhead{HST(11)\tablenotemark{b}}& \colhead{HIP97\tablenotemark{c}} & \colhead{HIP07\tablenotemark{d}} } 
\startdata
310&15.40$\pm$0.9&20.13$\pm$0.17&19.35$\pm$1.79&19.31$\pm$1.93\\
383&16.00$\pm$0.9&21.53$\pm$0.20& &\\
472&16.60$\pm$1.6&21.70$\pm$0.15&21.29$\pm$1.91&21.86$\pm$1.71\\
548&16.80$\pm$0.3&20.69$\pm$0.17& &\\
622&21.60$\pm$1.1&24.11$\pm$0.30& &\\
627&16.50$\pm$0.9&21.74$\pm$0.25&23.41$\pm$1.65&22.75$\pm$1.22\\
645&15.70$\pm$1.2&17.46$\pm$0.21&15.11$\pm$4.75&19.13$\pm$5.45\\
\\
weighted average&16.80$\pm$0.24&20.81$\pm$0.19&21.19$\pm$1.00&21.72$\pm$0.87& &\\
average&16.94$\pm$0.99&21.05$\pm$0.21&19.79$\pm$ 2.04&20.76$\pm$1.05& &\\
median&16.50$\pm$0.99&21.52$\pm$0.20&20.32$\pm$ 2.04&20.59$\pm$1.05 &\\
$average_{HIP}$\tablenotemark{e}&16.05$\pm$1.15&20.26$\pm$0.19&19.79$\pm$2.04&20.76$\pm$105\\
\\
\hline
\\
HIP mean parallax\tablenotemark{f}&&&&21.53$\pm$2.76$\pm$0.23& &\\
\enddata
\tablenotetext{a}{Parallaxes from the original analysis \citep{WvA97a}}
\tablenotetext{b}{Parallaxes from the present study (Section \ref{AST})}
\tablenotetext{c}{Parallaxes from the original \HIPs reduction \citep{Per97}}
\tablenotetext{d}{Parallaxes from the  \HIPs re-reduction \citep{Lee07a}}
\tablenotetext{e}{Parallax average considering only those stars in common with \HIP}
\tablenotetext{f}{The mean parallax of  selected \HIP ~~stars \citep{Lee09} }

\end{deluxetable}

\begin{deluxetable}{crrcc}
\tablewidth{0in}
\tablecaption{\HST Parallaxes in Common with \HIP\label{tbl-picomp}}
\tablehead{\colhead{ID}&\colhead{$\pi_{\HST}$\tablenotemark{a}}&
\colhead{$\pi_{\HIP}$\tablenotemark{b}}& \colhead{$\Delta \HIP$ \tablenotemark{c}} & \colhead{$\Delta \HST$\tablenotemark{c}} } 
\startdata
Proxima Cen\tablenotemark{d,e}&769.7$\pm$0.3&771.64$\pm$2.6&3.17&-0.04\\
Barnard's Star\tablenotemark{d,e}&545.5$\pm$0.3&548.31$\pm$1.51&3.49&-0.14\\
Feige 24\tablenotemark{e}&14.6$\pm$0.4&10.86$\pm$3.94&-3.97&0.04\\
Wolf 1062 AB\tablenotemark{d,g}&98$\pm$0.4&97.85$\pm$2.95&-0.25&0.00\\
RR Lyr\tablenotemark{h}&3.82$\pm$0.2&3.46$\pm$0.64&-0.60&0.06\\
delta Cep\tablenotemark{i}&3.66$\pm$0.15&3.81$\pm$0.2&-0.09&0.05\\
HD 213307\tablenotemark{i}&3.65$\pm$0.15&3.69$\pm$0.46&-0.23&0.02\\
GJ 876\tablenotemark{j}&214.6$\pm$0.2&213.26$\pm$2.12&-1.19&0.01\\
55 Cnc\tablenotemark{k}&79.78$\pm$0.3&80.55$\pm$0.7&0.53&-0.10\\
$\epsilon$ Eri\tablenotemark{l}&311.37$\pm$0.11&310.95$\pm$0.16&-0.06&0.03\\
HD 33636\tablenotemark{m}&35.6$\pm$0.2&35.27$\pm$1.02&-0.54&0.02\\
HD 136118\tablenotemark{n}&19.35$\pm$0.17&21.48$\pm$0.55&1.70&-0.16\\
l Car\tablenotemark{o}&2.01$\pm$0.2&2.06$\pm$0.27&-0.16&0.09\\
$\zeta$ Gem\tablenotemark{o}&2.78$\pm$0.18&2.71$\pm$0.17&-0.17&0.19\\
$\beta$ Dor\tablenotemark{o}&3.14$\pm$0.16&3.64$\pm$0.28&0.15&-0.05\\
W Sgr\tablenotemark{o}&2.28$\pm$0.2&2.59$\pm$0.75&0.01&0.00\\
X Sgr\tablenotemark{o}&3.00$\pm$0.18&3.39$\pm$0.21&0.05&-0.04\\
Y Sgr\tablenotemark{o}&2.13$\pm$0.29&3.73$\pm$0.32&0.72&-0.59\\
FF Aql\tablenotemark{o}&2.81$\pm$0.18&2.05$\pm$0.34&-0.83&0.23\\
T Vul\tablenotemark{o}&1.9$\pm$0.23&2.31$\pm$0.29&0.07&-0.04\\
RT Aur\tablenotemark{o}&2.4$\pm$0.19&-0.23$\pm$1.01&-2.83&0.10\\
HD38529\tablenotemark{p}&25.11$\pm$0.19&25.46$\pm$0.4&0.08&-0.02\\
$\upsilon$ And\tablenotemark{q}&73.71$\pm$0.10&74.25$\pm$0.72&0.37&-0.01\\
vA310\tablenotemark{r}&20.13$\pm$0.17&19.31$\pm$1.93&0.95&-0.01\\
vA472\tablenotemark{r}&21.70$\pm$0.15&21.86$\pm$1.71&3.07&-0.11\\
vA627\tablenotemark{r}&21.74$\pm$0.25&22.75$\pm$1.22&0.03&-0.01\\
vA645\tablenotemark{r}&17.46$\pm$0.21&19.13$\pm$5.45&3.45&-0.01\\
\enddata
\tablenotetext{a}{Parallaxes in mas from \HST}
\tablenotetext{b}{Parallaxes in mas from \cite{Lee07a}}
\tablenotetext{c}{Residuals in mas from the impartial regression, \HIP ~vs \HST (Figure~\ref{fig-HH})}
\tablenotetext{d}{Parallaxes resulting from analysis that used a Galaxy model-based correction to absolute parallax (YPC95) }
\tablenotetext{e}{Parallax from \cite{Ben99} }
\tablenotetext{f}{Parallax from \cite{Ben00a}}
\tablenotetext{g}{Parallax from \cite{Ben01}}
\tablenotetext{h}{Parallax from \cite{Ben02a}}
\tablenotetext{i}{Parallax from \cite{Ben02b}}
\tablenotetext{j}{Parallax from \cite{Ben02c}}
\tablenotetext{k}{Parallax from \cite{McA04}}
\tablenotetext{l}{Parallax from \cite{Ben06}}
\tablenotetext{m}{Parallax from \cite{Bea07}}
\tablenotetext{n}{Parallax from \cite{Mar10}}
\tablenotetext{o}{Parallax from \cite{Ben07b}}
\tablenotetext{p}{Parallax from \cite{Ben10}}
\tablenotetext{q}{Parallax from \cite{McA10}}
\tablenotetext{r}{Parallax from this paper}
\end{deluxetable}
\begin{deluxetable}{llll}
\tablewidth{0in}
\tablecaption{Hyad Distance Moduli and Absolute Magnitudes \label{tbl-AbsM}}
\tablehead{\colhead{vA}& \colhead{V}&
\colhead{m-M}& \colhead{M$_{\rm V}$}  } 
\startdata
310&10.01&3.43&6.53$\pm$0.018\\
383&12.20&3.35&8.86$\pm$0.023\\
472&9.03&3.32&5.71$\pm$0.015\\
548&10.33&3.39&6.91$\pm$0.018\\
622&11.90&3.09&8.81$\pm$0.027\\
627&9.55&3.31&6.24$\pm$0.025\\
645&11.00&3.79&7.21$\pm$0.026\\
\enddata
\end{deluxetable}

 \begin{center}
\begin{deluxetable}{lcl}
\tablewidth{0in}
\tablecaption{ Hyades Distance Moduli \label{tbl-mM}}
\tablehead{\colhead{m-M}&
\colhead{Author}& \colhead{Method } }
\startdata
3.28 $\pm$0.10&   \cite{Gun88}& RV \& Convergent Point\\
3.45&   \cite{Mor92} &Convergent Point\\
3.16$\pm$0.10   &\cite{Gat92}&Parallax of 51 Tauri\\
3.20$\pm$0.06&  \cite{Gat92}&Mean parallaxes (1992)\\
3.2$\pm$0.1 & \cite{Tur94} &Combined methods\\
3.40$\pm$0.07  &\cite{Tor97a} &Orbital parallax, 51 Tau \\
3.38$\pm$0.11  &\cite{Tor97b}  &Orbital parallax, 70 Tau \\
3.39$\pm$0.08   &\cite{Tor97c} &Orbital parallax, 78 Tau \\
3.42$\pm$0.09   &\cite{WvA97a} &\HST FGS (older reduction and modeling)\\
3.86$\pm$0.13   &\cite{WvA97a}&\HST FGS (unweighted average, all stars, Tables~\ref{tbl-pis}, \ref{tbl-AbsM})\\
3.32$\pm$0.06  & \cite{WvA97b}&Mean of ground-based parallaxes (YPC95)\\
3.33$\pm$0.01   & \cite{Per98}& \HIP\\ 
3.33$\pm$0.02   & \cite{Lee09} & \HIP~reanalysis\\
3.376$\pm$0.01   & this paper& \HST FGS reanalysis
\enddata
\end{deluxetable}
  \end{center} 

%
%
\clearpage

\begin{figure}[!h]
\centering
\epsscale{1.00}
\plotone{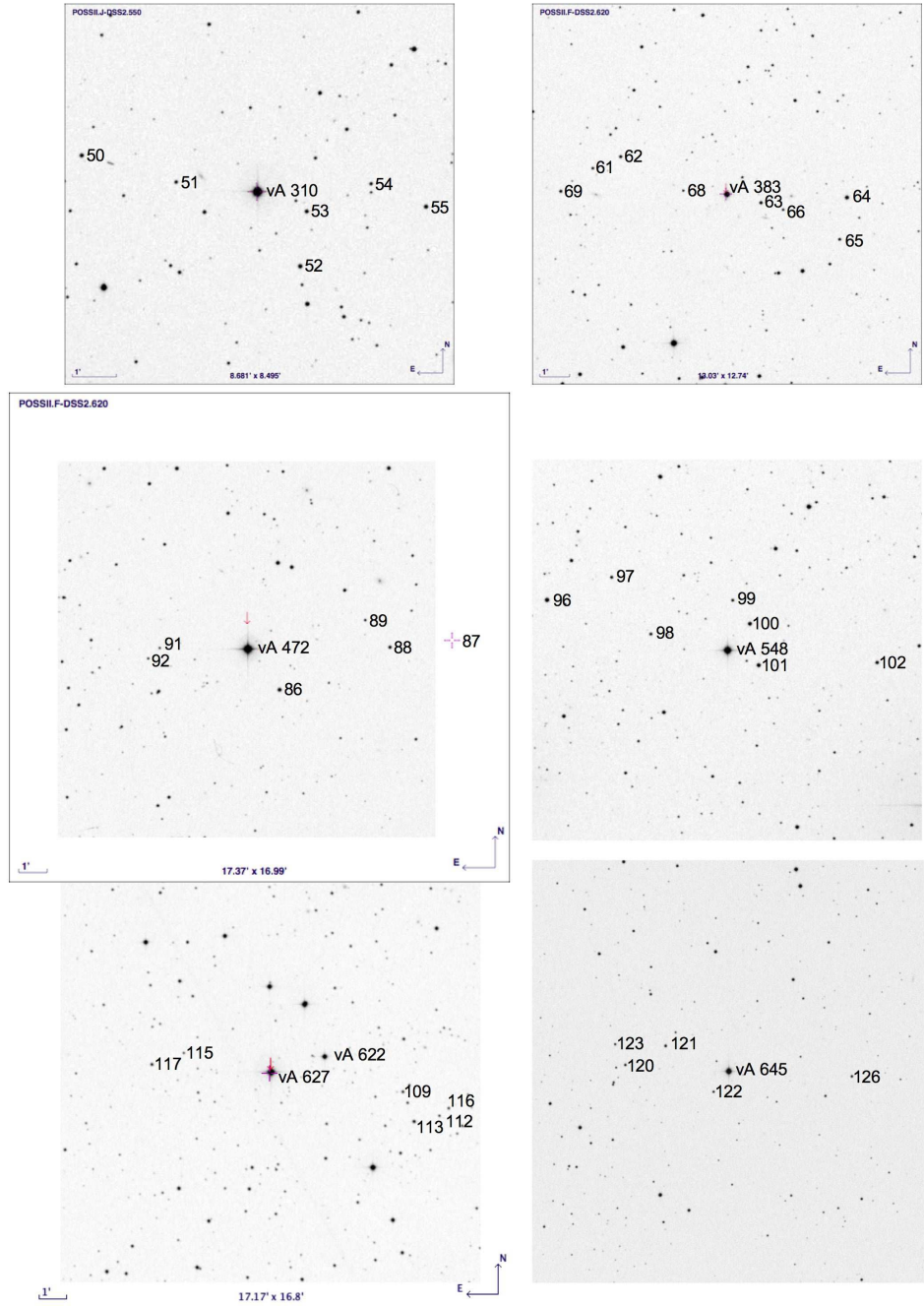}
\caption{Finding charts for subject Hyads  and astrometric reference stars.  Labels are immediately to the right of each star. Where scales are not indicated, the box size is $13\farcm1\times 13\farcm1$, north at top, east to left.}
\label{fig-1}
\end{figure}

\clearpage
\begin{figure}[!h]
\epsscale{0.75}
\plotone{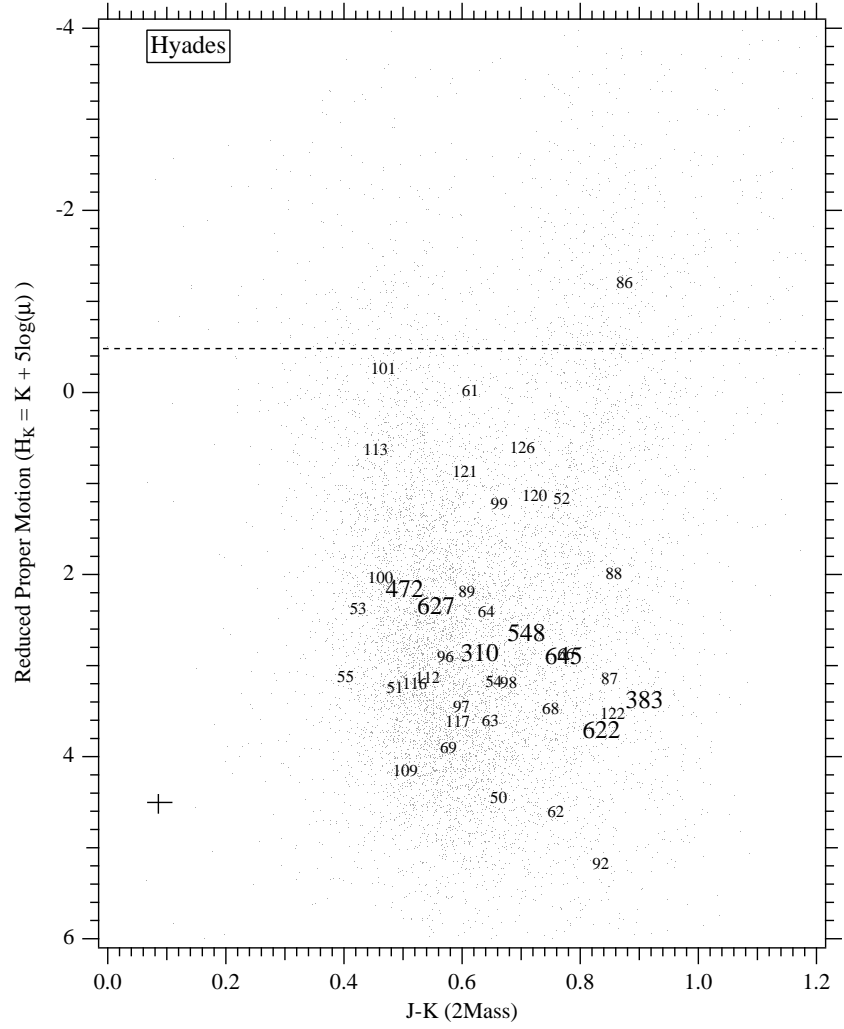}
\caption{Reduced proper motion diagram for 10,000 stars in a 2\arcdeg~field centered on the Hyades. Star identifications are van Altena numbers and our internal ID for astrometric reference stars (Table~\protect\ref{tbl-SPP}). H$_K$ for these stars is calculated using our final proper motions (Table~\protect\ref{tbl-POS}). For a given spectral type giants and sub-giants have more negative H$_K$ values and are redder than dwarfs in J-K.  The horizontal line is an estimated demarkation between dwarfs (below) and giants (above) from a statistical analysis of the Tycho input catalog (D. Ciardi 2004, private communication). The cross in the lower left corner indicates representative errors along each axis.}
\label{fig-RPM}
\end{figure}
\clearpage

\begin{figure}[!h]
\centering
\epsscale{1.00}
\plotone{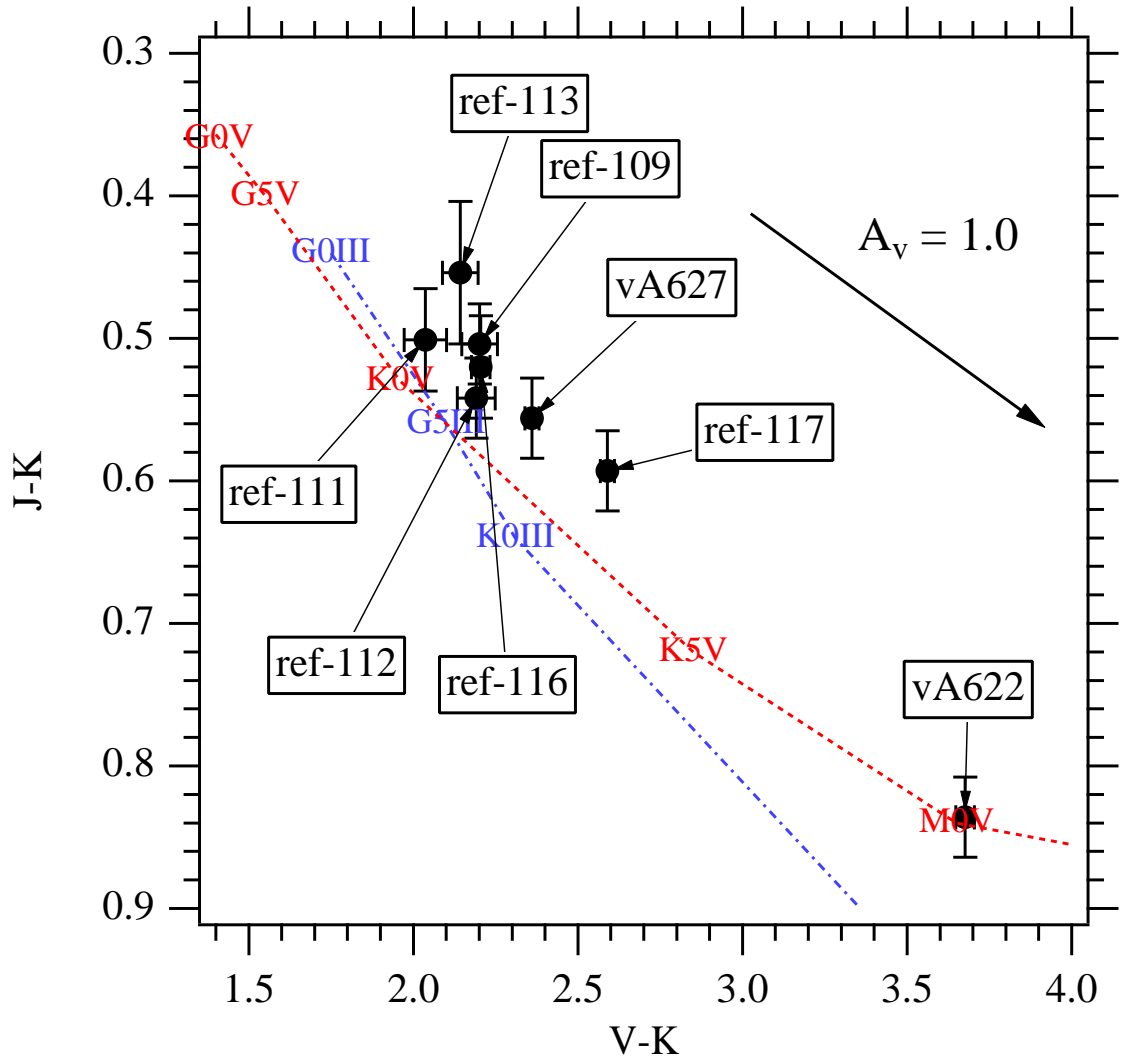}
\caption{ J-K vs V-K color-color diagram for the vA\,622  vA\,627 field. The dashed line is the locus of  dwarf (luminosity class V) stars of various spectral types; the dot-dashed line is for giants (luminosity class III). The reddening vector indicates A$_V$=1.0 for the plotted color system.}
\label{fig-CCD}
\end{figure}

\begin{figure}[!h]
\centering
\epsscale{0.65}
\plotone{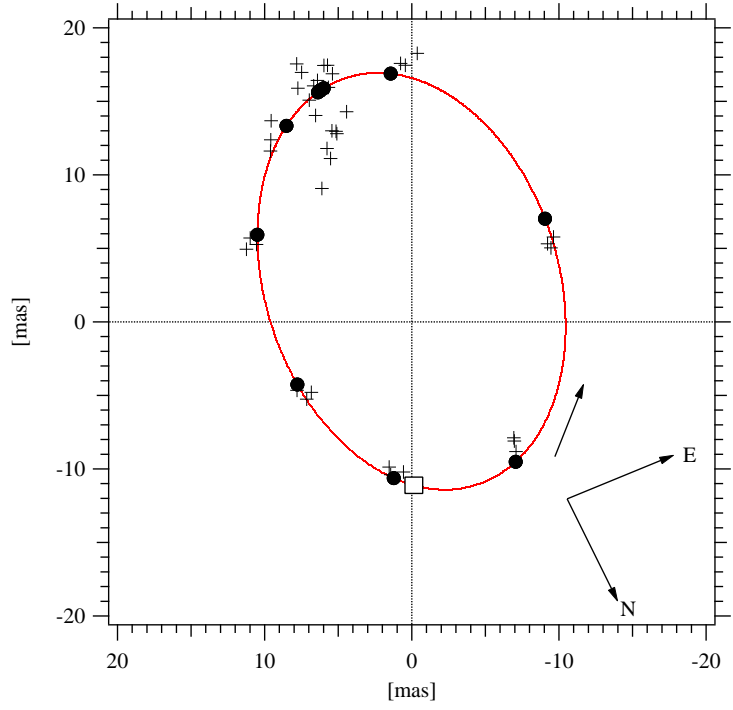}
\caption{ Final orbit and residuals for  the vA\,627 perturbation. The epoch of periastron passage is plotted as a square.} 
\label{orb}
\end{figure}
\clearpage

\begin{figure}[!h]
\centering
\epsscale{0.65}
\plotone{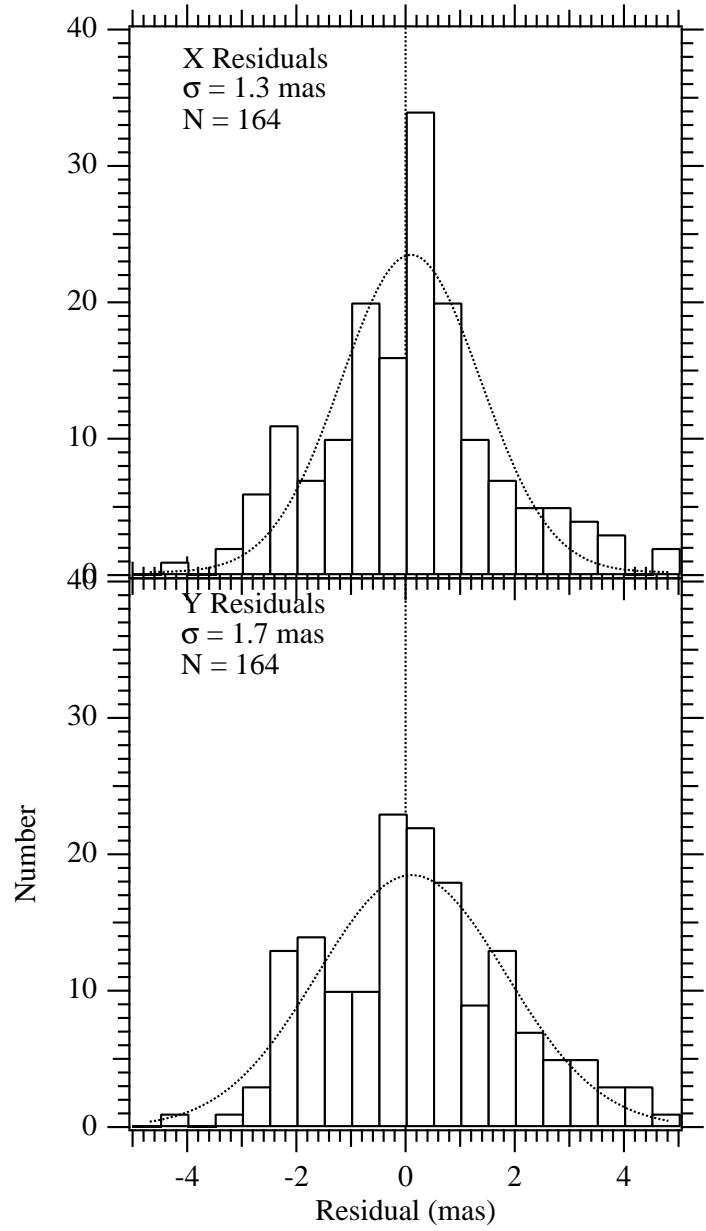}
\caption{ Histograms of x and y residuals obtained from modeling vA\,622,  vA\,627, and the astrometric reference stars with Equations 4 and 5, including Keplerian orbit terms for  vA\,627. Distributions are fit
with Gaussians whose $\sigma$'s are noted in the plots.} 
\label{fig-4}
\end{figure}
\clearpage

\begin{figure}[!h]
\centering
\epsscale{0.65}
\plotone{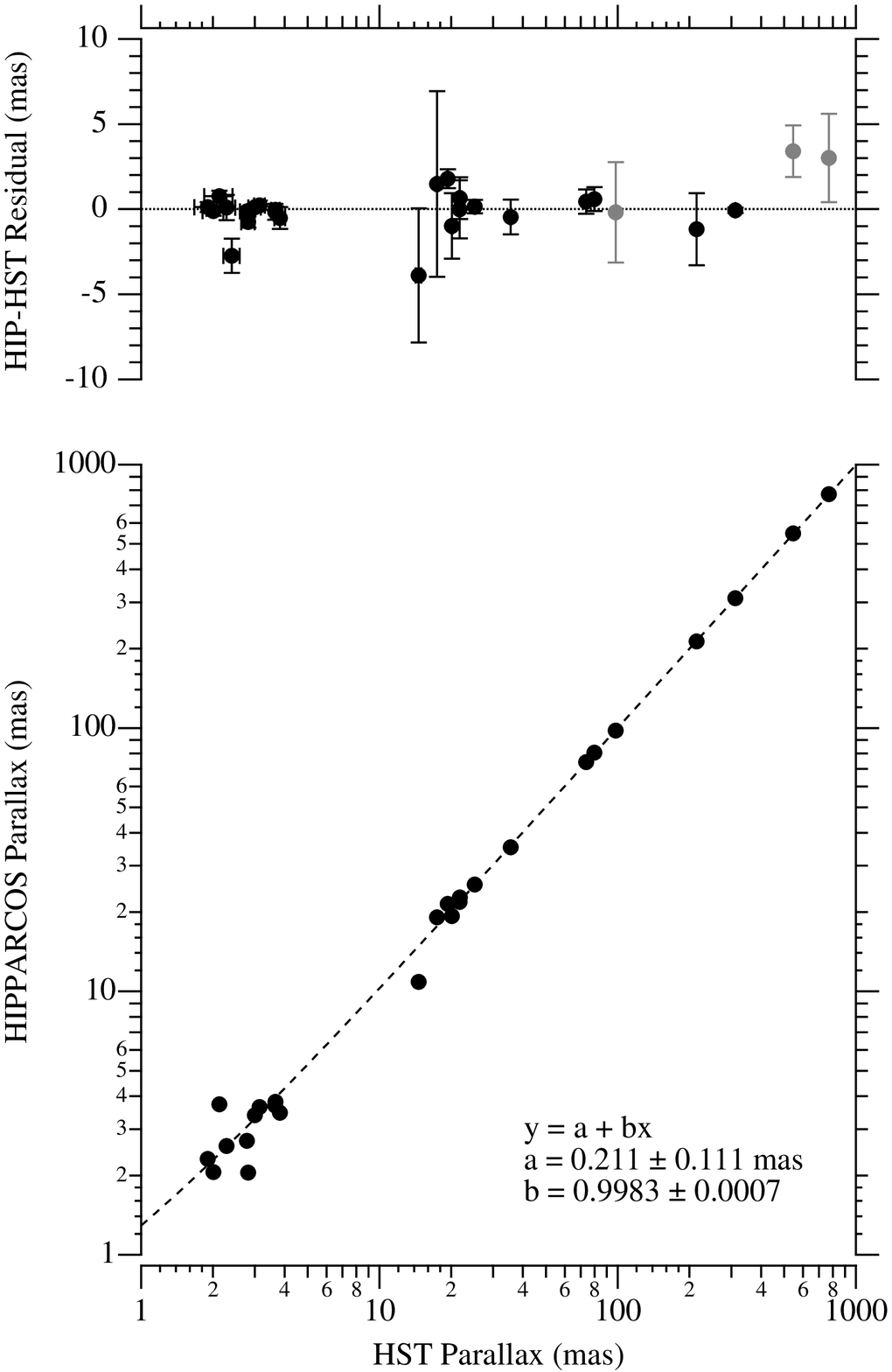}
\caption{ \HIP ~parallaxes from the re-reduction of \protect\cite{Lee07a} compared to all parallaxes from \HST FGS (Table~\protect\ref{tbl-picomp}). The regression line is impartial, in that errors in both \HIP~ and \HST parallaxes are considered \protect\citep{Jef88}. The few objects measured earlier (Proxima Cen, Barnard's Star, Wolf 1062 AB), employing a model-based correction to absolute parallax are plotted in lighter grey. Notable outliers include Feige 24 (faint CV), FF Aql, vA 645, and  Y Sgr. 
}
\label{fig-HH}
\end{figure}
\clearpage

\begin{figure}[!h]
\centering
\epsscale{0.65}
\plotone{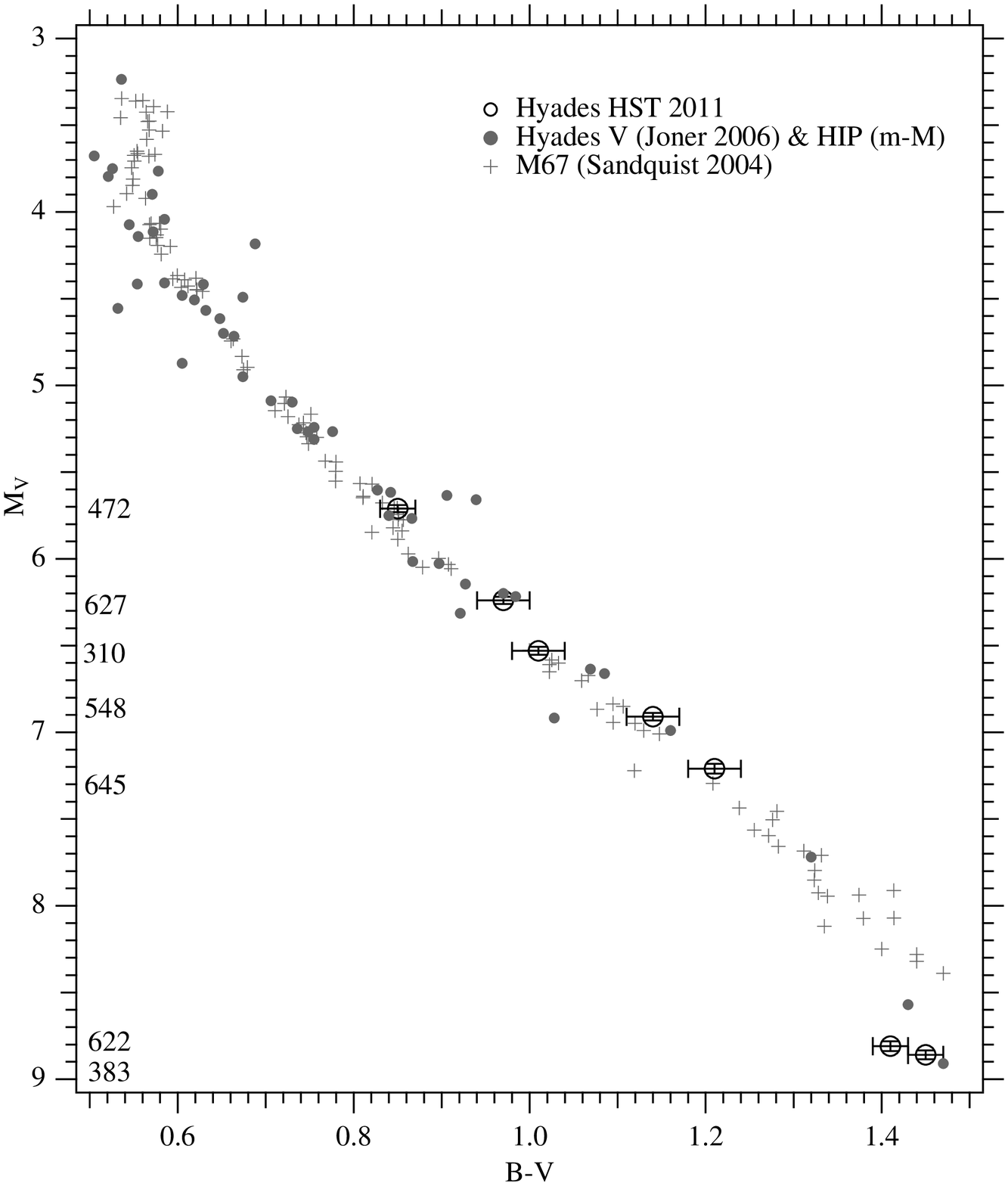}
\caption{ Absolute magnitude HR diagram for the seven Hyads with \HST parallaxes. These are identified by van Altena number just interior to the left axis. Parallaxes used are from Table~\protect\ref{tbl-pis}, `HST11', the results from Section~\protect\ref{PIS}. Error bars are $\pm1\sigma$. Also shown are M$_{\rm V}$ for the Hyads with photometry from \cite{Jon06}, assuming the  \protect\cite{Lee09} distance modulus, m-M=3.33.  Also plotted are M67 M$_{\rm V}$ from \protect\cite{San04}. vA\,627 is a known spectroscopic, now astrometric, binary. The  \HST parallaxes yield a weighted average distance modulus m-M=3.376 $\pm$ 0.01.} 
\label{fig-HR}
\end{figure}
\clearpage

\end{document}